\documentclass{INTERSPEECH2023}
\interspeechcameraready 
\usepackage{multirow}
\title{Implicit spoken language diarization}
\name{Jagabandhu Mishra, Amartya Chowdhury, S. R. Mahadeva Prasanna}
\address{
 Department of Electrical Engineering, Indian Institute of Technology (IIT) Dharwad, India
  }
  \email{jagabandhu.mishra.18, amartya.chowdhury, prasanna\}@iitdh.ac.in }

\begin{document}

\maketitle
\begin{abstract}
Spoken language diarization (LD) and related tasks are mostly explored using the phonotactic approach. Phonotactic approaches mostly use explicit way of language modeling, hence requiring intermediate phoneme modeling and transcribed data. Alternatively, the ability of deep learning approaches to model temporal dynamics may help for the implicit modeling of language information through deep embedding vectors. Hence this work initially explores the available speaker diarization frameworks that capture speaker information implicitly to perform LD tasks. The performance of the LD system on synthetic code-switch data using the end-to-end x-vector approach is $6.78\%$ and $7.06\%$, and for practical data is $22.50\%$ and $60.38\%$, in terms of diarization error rate and Jaccard error rate (JER), respectively. The performance degradation is due to the data imbalance and resolved to some extent by using pre-trained wave2vec embeddings that provide a relative improvement of $30.74\%$ in terms of JER.

\end{abstract}
\noindent\textbf{Index Terms}: Spoken language diarization. wav2vec, Jaccard error rate (JER), Acoustic similarity, Data imbalance  

\section{Introduction}
 Spoken language diarization (LD) is a task to automatically segment and label the monolingual segments present in a code-switched (CS) utterance. According to the humans' language abstraction level, acoustic-phonetic and phonotactic information are largely used in literature for the modeling of language-specific information~\cite{li2013spoken,V2018}. The acoustic-phonetic information mostly captures the information related to phoneme production, whereas the phonotactic information captures the phonemic distribution of the language~\cite{li2013spoken}. It is evident from the literature that phonotactic information better captures language-specific evidence than the acoustic-phonetic approach~\cite{lyu2013language,lyu2013language1}. However, most of the available phonotactic information-based frameworks require transcribed speech data, which makes the usability limited for resource-scare languages~\cite{surveycodeswitch2019}. Alternatively, the language information can be modeled in two ways: (a) implicit, and (b) explicit~\cite{nagarajan2004implicit}. Implicit approaches model the language information directly from the speech signals. In contrast, the explicit approaches, include the modeling of language information through intermediate representations like phonemes, Senones and tokens, etc~\cite{nagarajan2004implicit,vuddagiri2018iiith}. 

 Specific to LD, code-switched utterances are mostly uttered by a single speaker. In such a scenario, the phoneme production of  secondary language may be biased towards the primary, hence making language discrimination difficult at the acoustic-phonetic level. Therefore, most LD frameworks use phonotactic approaches to capture language-specific information~\cite{mishra2022issues,mishra2022importance}. In CS utterances, mostly either of the languages is a resource-scare in nature.  In such a scenario, it may be difficult to get the transcribed speech data to train the automatic speech recognition (ASR) system for deriving the phoneme distribution. Hence there is a need to explore alternative approaches for the development of LD. 

\begin{figure}
 \centering
\includegraphics[height= 125pt,width= 220pt]{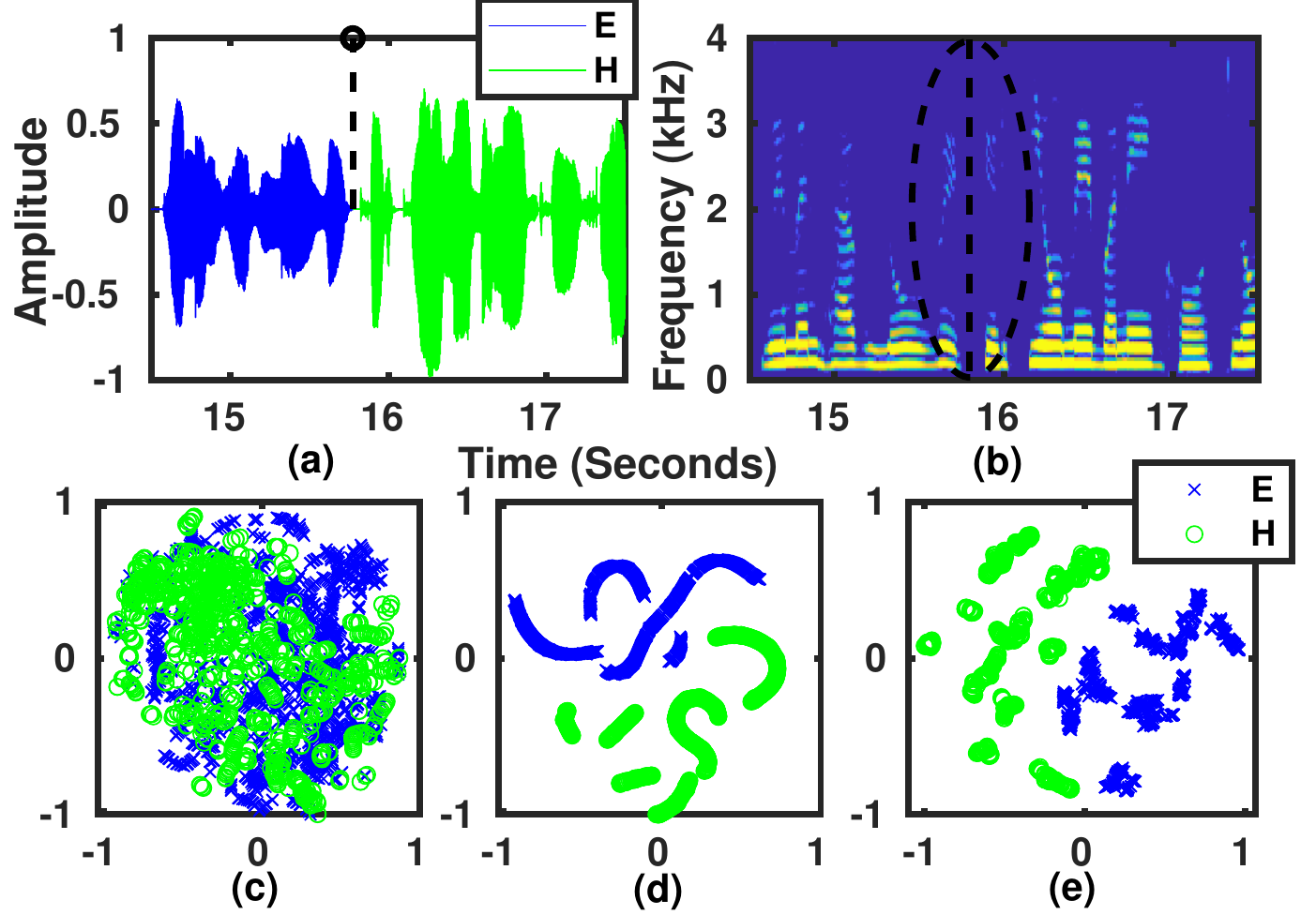}
 \caption{(a) Time domain representation of a Code-switched speech utterance, (b) spectrogram, (c) t-SNE distribution of the MFCC features, (d) i-vector and (e) x-vector representations, respectively.}
\vspace{-0.7 cm}
 \label{f1}
 \end{figure}

 The aim here is to capture better language-specific information using implicit way of language modeling. Hence the need is to capture the phonemic distribution and the ways they combined to form syllables and subwords etc. through implicit modeling. It  means that there is a requirement for the modeling of underlying long-term spectro-temporal dynamics.  Recently machine learning and deep learning (ML/DL) methods contribute to the development of i/x-vector-based approaches.  Generally, i/x-vector-based approaches mostly model the long-term spectro-temporal dynamics~\cite{vuddagiri2018iiith,snyder2018spoken}. Figure~\ref{f1}, shows the CS utterances, corresponding spectrogram,  t-SNE distribution of the MFCC feature, i-vector, and x-vector distribution of a CS utterance. From the figure, it can be observed that language discrimination is difficult directly from the time domain signal and its spectrogram. As hypothesized, due to the bias of phoneme production, it is difficult to discriminate at the feature level. However, the statistical i-vectors and time delay neural network-based (TDNN) x-vectors are known for capturing long-term spectro-temporal dynamics, hence showing a better cluster between the languages in Figure~\ref{f1} (d) and (e), respectively. This motivates the development of LD frameworks through implicit approaches.
 
Speaker diarization (SD), a task similar to LD, is well explored in the literature. Fortunately, most of the frameworks available in the SD literature follow the implicit way of speaker modeling. Mostly the available SD frameworks can be broadly classified into three approaches: (a) change point, (b) clustering, and (c) end-to-end-based approach. Hence as an initial attempt, this work plans to perform LD using the available SD frameworks.

 The rest of the paper is organized as follows: Section~\ref{mexs} describes the brief review of LD and related works. In Section~\ref{Imp} the implicit approaches for SD and LD are described. The experimental setup and results are discussed in Section~\ref{exp}. Finally, the conclusion and future directions are discussed in Section~\ref{conclusion}.

\vspace{-0.20 cm} 
\section{Review of Spoken Language Diarization}
\vspace{-0.10 cm} 
 \label{mexs}
The attempts towards LD are limited in the literature. However, there exists some work, that performs code-switch detection (CSD), code-switch utterance detection (CSUD), sub-utterance language identification (SLID), etc~\cite{shah2020first,lyu2013language,mishra2022issues}. The summary of the attempted approaches is tabulated in Table~\ref{rv}. 

From the table, it can be observed that most of the attempts try to model language, either explicitly or implicitly by capturing the phoneme distribution and flow to form syllables and words. In ~\cite{lyu2013language,lyu2013language1} and ~\cite{barras2020vocapia}, the work uses phoneme sequence (PS) derived from the n-gram model, Gaussian mixture model (GMM) posterior, and i-vector for performing SLID task. The PS uses explicit and the i-vector and GP uses the implicit approach for language modeling. The works in ~\cite{V2018} and ~\cite{yilmaz2017language} use bottleneck features, extracted from the trained Senone model and further use the i-vector framework to capture language-specific information. After the evolution of deep learning approaches, in ~\cite{shah2020first,liu2021end} and ~\cite{krishna2020utterance}, the works used deepspeech2 (DS2), transformer, x-vector based frameworks to implicitly model the language information and perform end-to-end tasks like SLID and CSUD.  

In~\cite{lyu2013language,lyu2013language1} and ~\cite{barras2020vocapia}, the work concludes that for the CS scenarios to capture language-specific evidence explicit modeling is preferable over implicit. However, the performance achieved for the CSUD task using implicit modeling in ~\cite{liu2021end} and explicit modeling in~\cite{barras2020vocapia} is comparable. The advantages of the implicit approach over the explicit approach are: (a) it doesn't rely on the performance of intermediate modeling, and (b) it doesn't require transcribed speech data. Therefore, these approaches can be easily adapted for low-resource and resource-scarce languages. Hence motivated to explore implicit approaches to perform LD tasks.

\begin{table}[]
\tiny
\centering
\caption{Summary of related works, CER: character error rate, WER: word error rate, LF: latent feature, PS: phoneme sequence, DS2: Deepspeech2, CRF: conditional random field. }
\label{rv}
\begin{tabular}{|l|ll|l|l|l|}
\hline
\textbf{System} &
 \multicolumn{2}{l|}{\textbf{Methods}} &
 \textbf{Dataset} &
 \textbf{Task} &
 \textbf{Performance} \\ \hline
Lyu et.al~\cite{lyu2013language} &
 \multicolumn{2}{l|}{\begin{tabular}[c]{@{}l@{}}PS\\ CRF\end{tabular}} &
 SEAME &
 SLID &
 FER:14.4 \\ \hline
Lyu et.al~\cite{lyu2013language1} &
 \multicolumn{2}{l|}{\begin{tabular}[c]{@{}l@{}}PS, GP\\ CRF\end{tabular}} &
 SEAME &
 SLID &
 FER:14.7 \\ \hline
Yilmaz et.al~\cite{yilmaz2017language} &
 \multicolumn{2}{l|}{\begin{tabular}[c]{@{}l@{}}BNF\\ I-vector\end{tabular}} &
 FAME &
 ASR &
 \begin{tabular}[c]{@{}l@{}}WER: 12.7\\ (reduction)\end{tabular} \\ \hline
Spoothy et.al~\cite{V2018} &
 \multicolumn{2}{l|}{\begin{tabular}[c]{@{}l@{}}BNF\\ SVM\end{tabular}} &
 NGBC &
 CSD &
 IDA:86.16 \\ \hline
Sreeram et.al~\cite{sreeram2020joint} &
 \multicolumn{2}{l|}{\begin{tabular}[c]{@{}l@{}}Spectrogram\\ E2E attention\end{tabular}} &
 Hingcos &
 SLID &
 \begin{tabular}[c]{@{}l@{}}CER:23.47\\ WER:16.6\end{tabular} \\ \hline
Shah et.al~\cite{shah2020first} &
 \multicolumn{2}{l|}{\begin{tabular}[c]{@{}l@{}}Spectrogram\\ DS2\end{tabular}} &
 MSCS &
 \begin{tabular}[c]{@{}l@{}}CSUD (A)\\ SLID (B)\end{tabular} &
 \begin{tabular}[c]{@{}l@{}}IDA (A):74\\ IDA (B):76.9\end{tabular} \\ \hline
Rangan et.al~\cite{rangan2020exploiting} &
 \multicolumn{2}{l|}{\begin{tabular}[c]{@{}l@{}}Spectrogram\\ DS2+L2-mask\end{tabular}} &
 MSCS &
 \begin{tabular}[c]{@{}l@{}}CSUD (A)\\ SLID (B)\end{tabular} &
 \begin{tabular}[c]{@{}l@{}}IDA (A):76.8\\ IDA (B):76.2\end{tabular} \\ \hline
Gauvain et.al~\cite{barras2020vocapia} &
 \multicolumn{2}{l|}{\begin{tabular}[c]{@{}l@{}}MFCC\\ PS, I-vector\end{tabular}} &
 MSCS &
 \begin{tabular}[c]{@{}l@{}}CSUD (A)\\ SLID (B)\end{tabular} &
 \begin{tabular}[c]{@{}l@{}}IDA (A):83.3\\ IDA (B):81.2\end{tabular} \\ \hline
Rallabandi et.al~\cite{rallabandi2020detecting} &
 \multicolumn{2}{l|}{\begin{tabular}[c]{@{}l@{}}LF\\ VB encoder\end{tabular}} &
 MSCS &
 CSUD &
 IDA:76.1 \\ \hline
Krishna et.al~\cite{krishna2020utterance} &
 \multicolumn{2}{l|}{\begin{tabular}[c]{@{}l@{}}Spectrogram\\ Transformer\end{tabular}} &
 MSCS &
 CSUD &
 IDA:79.82 \\ \hline
Liu et.al~\cite{liu2021end} &
 \multicolumn{2}{l|}{\begin{tabular}[c]{@{}l@{}}X-vector\\ Deep Clustering\end{tabular}} &
 MSCS &
 SLID &
 IDA:82.56 \\ \hline
\end{tabular}
\end{table}

\begin{figure}[hbt!]
    \includegraphics[height= 140pt,width= 225pt]{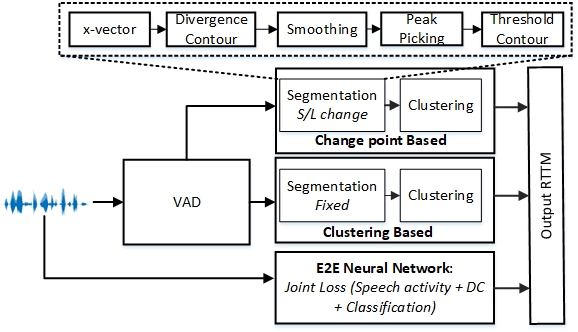}
    \caption{Block diagram depicting implicit approaches for SD/LD, DC: Deep clustering, VAD: Voice activity detector.}
    \label{iap}
\end{figure}

\vspace{-0.20 cm}
\section{Implicit Approaches for Speaker and Language Diarization}
\vspace{-0.15 cm}
\label{Imp}
The SD frameworks that use an implicit approach to model speaker information are broadly classified into three groups: (a) change point-based approach, (b) clustering-based approach, and (c) end-to-end based approach. A summary block diagram of the approaches is depicted in Figure~\ref{iap} and detailed descriptions of each approach are described in the following subsections.
\subsection{Change point-based approach}
The change detection framework used here is inspired by the speaker change detection framework available at~\cite{moattar2012review,tranter2006overview,park2022review}. Initially, the speech signal is passed through a short-term energy (STE) based voice activity detector, and the voiced frame locations are stored. The MFCC features are extracted from each utterance and taking reference from the voiced frames, the features belonging to the voiced frames are used for further processing. The voiced feature vectors belonging to each analysis window are used to extract the x-vectors. The divergence distance is the probabilistic linear discriminate analysis (PLDA) distance between the two consecutive x-vectors. The same setup with an analysis window length of $N$ and a shift of $1$ frame is used to compute the divergence contour. The contour is smoothed using a hamming window, with a length of $w_{l}$. The peak picking with a minimum distance parameter $\gamma$ detects the peak on the smoothed divergence contour. The final peak locations are decided by comparing the peak strength with the threshold contour used in~\cite{lu2002speaker}. The final peak locations' corresponding voiced frame sample locations are decided as the change points. After change detection, for a given utterance, around the midpoint of each segment with $N$ feature vectors, x-vectors are extracted and clustered using agglomerative hierarchical clustering (AHC) with PLDA as a distance matrix. Further using the clustered labels the predicted  rich transcription time marked (RTTM) files are obtained for each test utterance. The stopping criteria of AHC is set to the maximum number of speakers/languages i.e. $2$. A detailed description of the change point approach can be found at~\cite{mishra2023spoken}.
\subsection{Clustering based approach}
Similar to the change point approach, the voiced feature vectors are decided using the STE-based VAD. After that instead of speaker/language-based segmentation, a fixed duration segmentation strategy is used with an analysis window length of $N$ and a shift of $1$ frame is used to segment the test utterances. From each segment, the x-vectors are extracted and clustered using AHC. After clustering, the clustered labels are used to obtain the predicted RTTM files for each test utterance.
\subsection{End-to-End based approach}
The end-to-end (E2E) framework used here is inspired by the LD study reported in~\cite{liu2021end}. The architecture is designed to view the diarization problem as a sub-utterance level classification task. The framework has two blocks: (1) x-vector-based classification, and (2) transformer-based deep clustering. Instead of using an initial VAD, the framework uses silence as a class along with the speakers/languages. The parameters of the architecture are trained using a joint loss of classification and clustering.  For each test utterance, the architecture will predict the sequence of labels. The sequence of labels is used to predict the RTTM file.
                                                                                                            

\vspace{-0.3 cm}
\section{Experimental Setup, Result, and Discussions}
\label{exp}
\subsection{Database setup}
In this work, we have used  synthetic CS data generated from the IIT Madras text-to-speech (IITM-TTS) corpus~\cite{baby2016resources}, and Microsoft code switch task-B (MSCS) corpus~\cite{shah2020first}.

The IITM-TTS corpus consists of recordings from a speaker in both the native and English languages. This work only considers a female speaker speaking Hindi and English to generate CS utterances. Similarly, an Assamese speaker speaking English, and a Hindi speaker speaking English are considered to generate multispeaker utterances. The detailed data generation is inspired by the work reported at~\cite{mishra2023spoken}. From the total duration, $5$ hours per language/speaker have been kept for training and the rest are used to generate $4000$ CS/multi-speaker utterances. The generated utterances have $1$-$5$ language/speaker change points. The average mono-lingual/mono-speaker segment duration of the generated utterances is approximately $5$ seconds. The generated dataset for SD and LD study is termed TTSF-SD and TTSF-LD, respectively.

The MSCS corpus is a practical dataset, consisting of conversational recordings in three language pairs: Gujarati-English (GUE), Tamil-English (TAE), and Telugu-English (TEE). The dataset has two partitions: training and development. The training and testing partition consists of CS utterances of approximately $16$ hours and $2$ hours for each language pair. The average monolingual segment duration for primary and secondary (English) language is approximately $1.5$
 and $0.5$ seconds, respectively.

\subsection{Performance Measure}
Mostly for SD tasks, the DER and JER are used as evaluation measures~\cite{moattar2012review,ryant2018first}. The evaluation measures that are used in LD literature are accuracy, equal error rate (EER), and frame error rate (FER)~\cite{lyu2013language,shah2020first,V2018}. However, it is observed from the available practical datasets that the duration of the primary language is comparatively much more than the secondary for a given utterance~\cite{mishra2022issues}. In the MSCS dataset, the ratio of primary and secondary language duration for each utterance has approximately $4:1$. Hence the use of accuracy, EER, and FER will provide biased performance toward the primary language. Similarly, if there exists a duration imbalance between the classes in the test utterances, the SD literature suggests the use of JER instead of DER, ~\cite{ryant2018first}. Therefore the JER is a better performance measure for evaluating the LD system performance. For comparison purposes, this study uses accuracy, EER, and DER along with JER to evaluate the performance of the LD systems.   

\subsection{Experimental setup}
The initial experiments are carried out on synthetic datasets using change point, clustering, and E2E approach. For all the approaches x-vectors are used as a representation. For the change point and clustering study, the $39$ dimensional MFCC feature vectors are extracted from speech signal with a framesize and frameshift of $20$ and $10$ msec respectively. For VAD, $6\%$ of the average frame energy of a given utterance is used to decide on the voiced/unvoiced frame. For LD and SD, $N$ is considered as  $200$, and $50$ disjointly to train the x-vector model, respectively. The value of $N$ is decided experimentally by observing validation loss and accuracy. The models for both speaker and language are trained for $20$ epochs. After that, observing the validation loss and accuracy, the models that belong to the $11^{th}$ and $15^{th}$ epoch are considered as an x-vector extractor for the SD and LD study, respectively. The x-vector implementation available in the speech brain is used here~\cite{ravanelli2021speechbrain}. For the language model, a dropout of $0.2$ is used in the second, third, fourth, and sixth layers along with L2 normalization. The speaker model is trained without using dropout and L2 normalization. 

During testing, for the change point and clustering approach, for both SD and LD tasks the analysis window shift, is considered as $1$ and the length  $N$ is considered as $50$ and $200$ respectively. For the change point-based approach, the speaker and language segments are obtained by considering ($\alpha$, $\delta$, and $\gamma$) as ($2.6$, $1.3$, and $0.9$) and ($3.2$, $1.3$, and $0.9$), respectively. The $\alpha$ is a hyper-parameter used to obtain the threshold contour and the hamming window length is $1/\delta$ time of $N$. The hyper-parameters  are decided by observing the change detection performance on the first $100$ test trails.

For, E2E based approach, the hyper-parameters and the feature dimensions (i.e $437$ for each $200$ msec duration) mentioned in ~\cite{liu2021end} are used here. The models are trained for $100$ epochs and the model provides the best validation accuracy used for testing. The models are trained with a learning rate of $0.001$. For the MSCS dataset, the models for each language pair are trained for $60$ epochs, with a learning rate of $0.001$.
\subsection{Results and discussion}
\begin{table}[hbt!]
\centering
\caption{Performance of SD and LD on the synthetic dataset, PM: performance measure, CP: change point, and CL: clustering approach.}
\begin{tabular}{|c|c|c|c|c|}
\hline
 & PM & CP & CL & E2E \\ \hline
\multirow{2}{*}{TTSF-SD} & DER & 6.84 & 10.03 & 5.17 \\ \cline{2-5} 
 & JER & 13.42 & 16.53 & 5.07 \\ \hline
\multirow{2}{*}{TTSF-LD} & DER & 11.16 & 18.56 & 6.78 \\ \cline{2-5} 
 & JER & 20.61 & 29.39 & 7.06 \\ \hline
\end{tabular}

\label{per_syn}
\end{table}

The results obtained on synthetic data for LD and SD study using change point, clustering, and E2E-based approach are tabulated in Table~\ref{per_syn}. For SD using the change point approach, the obtained performance in terms of DER and JER is $6.84$ and $13.42$, respectively. For, LD using the change point approach the obtained DER and JER are $11.16$ and $20.61$, respectively. The performance of SD and LD using the clustering-based approach is $10.03$ and $18.56$ in terms of DER,  $16.53$, and $29.39$ in terms of JER, respectively. The degradation of the performance from the change point to the clustering approach is due to not using any smoothing approach in the clustering-based approach. The advantage of smoothing is that it can smooth out sudden spikes and the disadvantage is it ended up with miss classifications in the boundary region. Further, it is difficult to decide upon the length of the smoothing window, if the distributions of mono-speaker/language segment duration have higher variance.

\begin{figure}
\centering
\includegraphics[height= 100pt,width= 200pt]{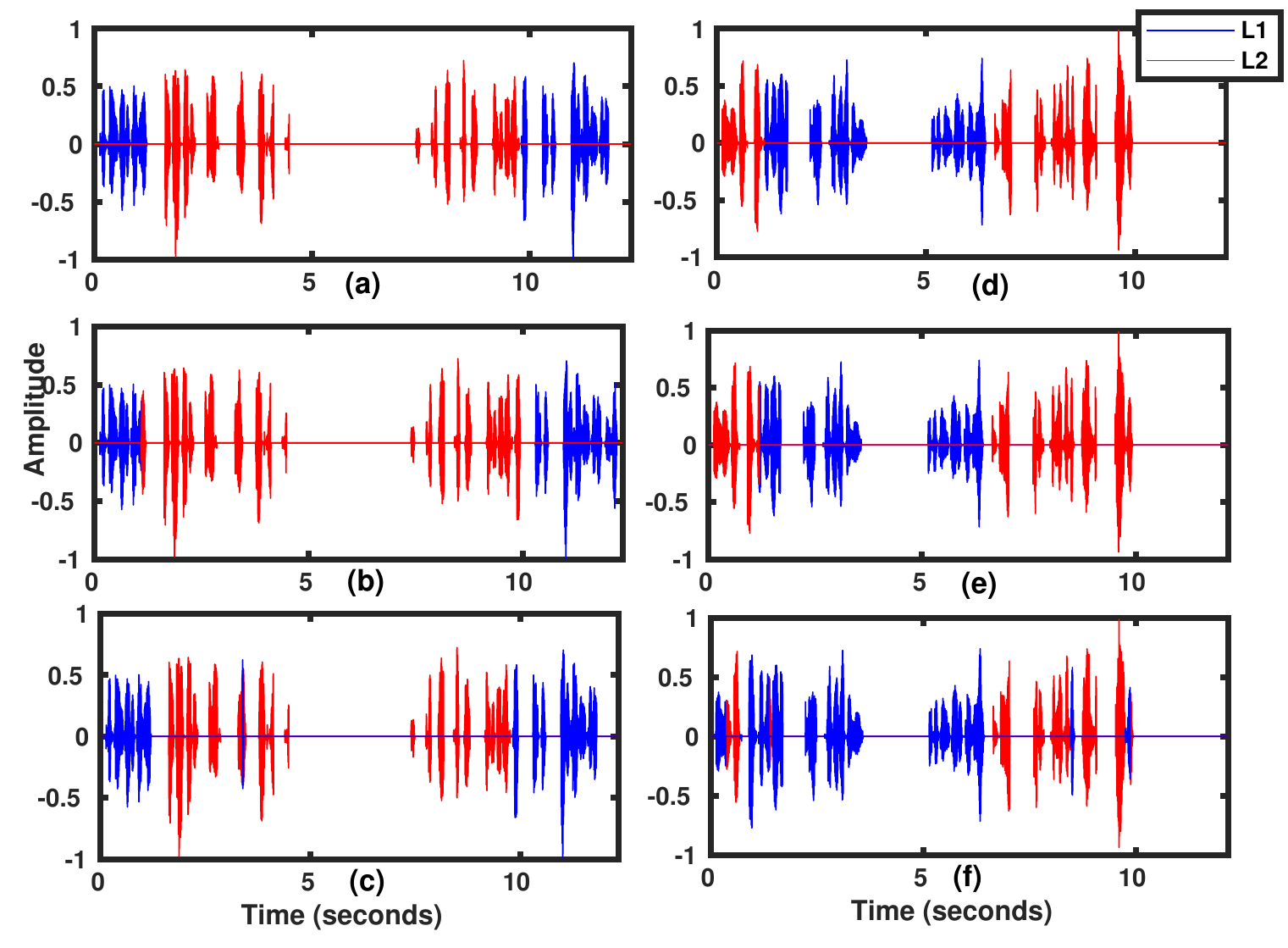}
\caption{(a),(d) CS speech, (b) and (e) extracted from change point approach (DER: $12.5$,$0.86$), and (c),(e) clustering approach (DER: $1.12$,$13.3$), L1 and L2 represent primary and secondary language.}
\vspace{-0.3 cm}
 \label{drawback_ch}
\end{figure}


 Figure~\ref{drawback_ch}(b), shows due to the boundary miss classification the DER is high, whereas in (c) using the clustering approach the DER is comparatively better. Similarly in Figure~\ref{drawback_ch}(f), due to sudden spikes, the DER is higher in the clustering-based approach as compared to the change point-based approach shown in Figure~\ref{drawback_ch}(e). Further, this shows both approaches are complementary to each other. After the evaluation of deep learning-based E2E frameworks, the joint classification and clustering loss were able to resolve the issue and improve the performance. Using the E2E framework, for SD the performance is $5.17$ and $5.07$, for LD the performance is $6.78$ and $7.06$ in terms of DER and JER, respectively. The performance of SD is comparatively better than the performance of LD. This is due to the ability of the x-vector to model the speaker is better than the language.

\begin{table}[hbt!]
\small
\centering
\caption{Performance of LD on MSCS dataset, Acc: Accuracy.}
\begin{tabular}{|l|lll|lll|}
\hline
 & \multicolumn{3}{l|}{x-vector (E2E)} & \multicolumn{3}{l|}{w2v embeddings} \\ \hline
 & \multicolumn{1}{l|}{GUE} & \multicolumn{1}{l|}{TAE} & TEE & \multicolumn{1}{l|}{GUE} & \multicolumn{1}{l|}{TAE} & TEE \\ \hline
DER & \multicolumn{1}{l|}{22.65} & \multicolumn{1}{l|}{22.86} & 22.01 & \multicolumn{1}{l|}{22.31} & \multicolumn{1}{l|}{25.83} & 21.75 \\ \hline
JER & \multicolumn{1}{l|}{60.55} & \multicolumn{1}{l|}{60.53} & 60.07 & \multicolumn{1}{l|}{40.51} & \multicolumn{1}{l|}{45.01} & 39.97 \\ \hline
Acc  & \multicolumn{1}{l|}{80.95} & \multicolumn{1}{l|}{81.48} & 81.75 & \multicolumn{1}{l|}{83.15} & \multicolumn{1}{l|}{79.05} & 82.35 \\ \hline
EER  & \multicolumn{1}{l|}{6.34} & \multicolumn{1}{l|}{6.45} & 6.08 & \multicolumn{1}{l|}{5.61} & \multicolumn{1}{l|}{6.98} & 5.88 \\ \hline
\end{tabular}

\label{mscs-per}
\end{table}

\begin{figure}[hbt!]
\centering
\includegraphics[height= 100pt,width= 200pt]{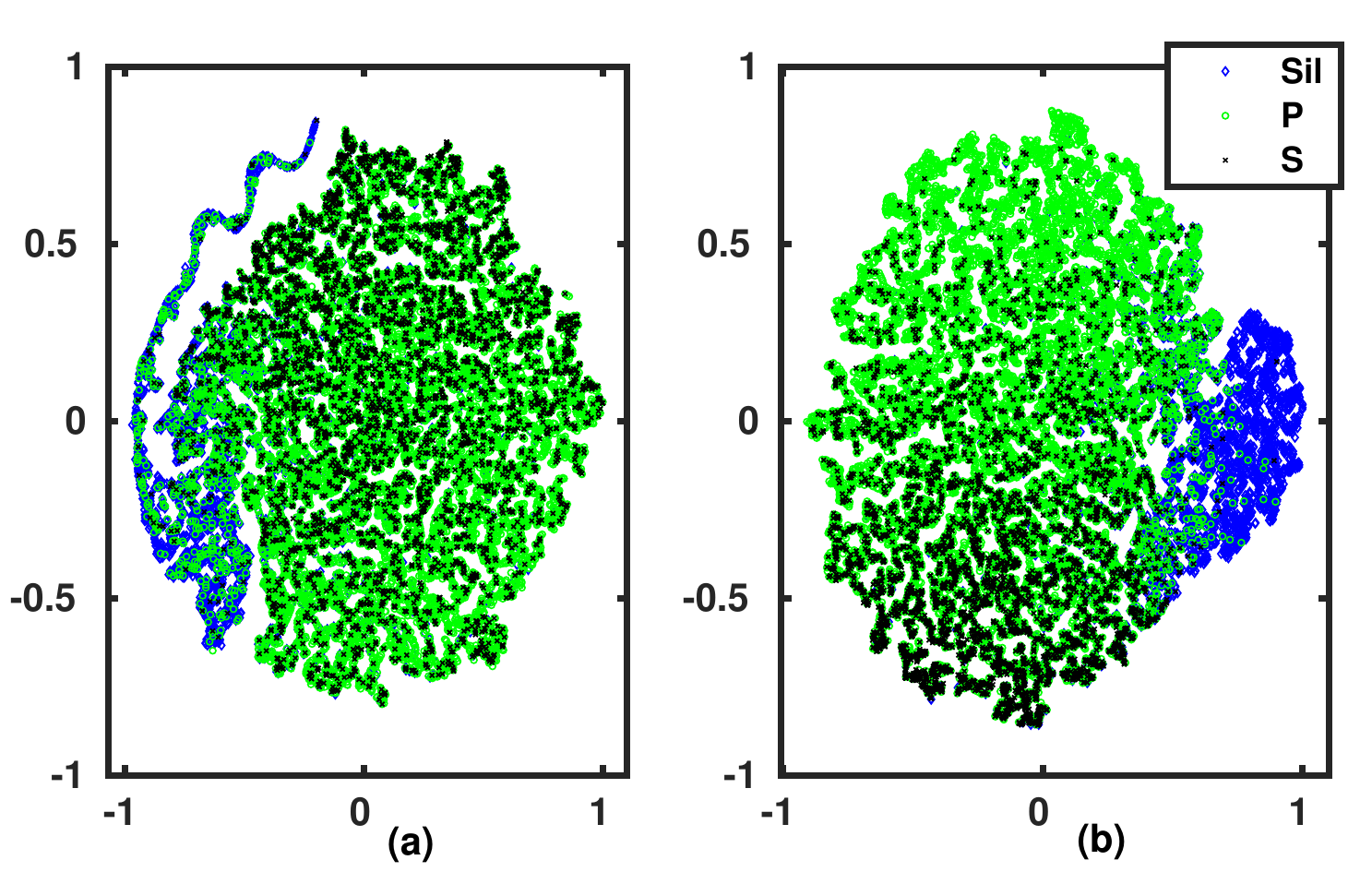}
\caption{t-SNE distribution (a) x-vector, (b) W2V based E2E framework.}
 \label{tsn_x_w2v}
\end{figure}

 The study is extended to a practical MSCS dataset with the x-vector-based E2E framework. The obtained performance in terms of DER, JER, accuracy, and EER is tabulated in Table~\ref{mscs-per}. It is observed that the performance in terms of Accuracy is $80.95\%$, $81.48\%$, and $81.75\%$, and in terms of JER is $60.55$, $60.53$, and $60.07$  for GUE, TAE, and TEE, respectively. Though the accuracy is around $80\%$ for all three language pairs, the difference in DER and JER values suggests the performance is biased towards one language. To validate the same, the confusion matrix is computed and tabulated in Table~\ref{conf}. From the table, it is observed that the performance is biased toward primary language, the same can also be observed from the t-SNE plot depicted in Figure~\ref{tsn_x_w2v}. The system is not predicting the secondary language. This is due to the unavailability of sufficient secondary language data to learn the discrimination between primary and secondary languages.

\begin{table}[hbt!]
\centering
\caption{Confusion Matrix, P: primary, S: secondary, and Sil: silence, respectively.}
\begin{tabular}{|l|l|l|l|l|}
\hline
 &  & P & S & Sil \\ \hline
\multirow{3}{*}{x-vector} & P & 90.64 & 0 & 9.36 \\ \cline{2-5} 
 & S & 65.23 & 0 & 34.77 \\ \cline{2-5} 
 & Sil & 16.83 & 0 & 83.17 \\ \hline
\multirow{3}{*}{w2v} & P & 86.05 & 4.44 & 9.51 \\ \cline{2-5} 
 & S & 21.97 & 54.56 & 23.47 \\ \cline{2-5} 
 & Sil & 8.68 & 8.34 & 82.98 \\ \hline
\end{tabular}

\label{conf}
\end{table}

 One way to resolve the issue is to use a pre-trained framework that has the ability to capture the language-specific long-term temporal dependence. Hence, this work uses a wav2vec-based (W2V) pre-trained architecture as a feature extractor that is trained using approximately $10000$ hours of speech utterances from $23$ Indian languages~\cite{gupta2021clsril}. The w2v pre-trained framework is trained using contrastive divergence loss to predict the embedding of the masked region of a given utterance. Hence, the hypothesis here is that the network may have captured the long-term temporal dependencies on Indian languages. The use of the same as a feature extractor instead of the x-vector extractor may improve the LD performance.

 The E2E-based framework is modified by taking W2V outputs passed through statistical pooling and two input layers of size $3000$ and $256$ to give input for the clustering block. The output of the two linear layers is passed again through two linear layers of size $256$ and $3$ to compute the classification loss. The linear layers except the last layer are used with batch normalization. The network is trained for $60$ epoch with a learning rate  of $0.001$. The obtained result is tabulated in Table~\ref{mscs-per}. The performance obtained in terms of JER is $40.51$, $45.01$, and $39.97$ for GUE, TAE, and TEE language pairs, respectively, and provides an average improvement of $30.74\%$. Further, the t-SNE distribution in Figure~\ref{tsn_x_w2v} and the confusion matrix in Table~\ref{conf} suggests the primary language bias is reduced to some extent.  
\vspace{-0.2 cm}
\section{Conclusion and Future work }
\label{conclusion}
In this study, the implicit approach is explored to perform the LD task. The performance of LD on synthetic data using change point, clustering, and E2E approach is comparable with the SD task. Extending to MSCS practical dataset, it is observed that the model output is biased toward the primary language. This is due to the unavailability of sufficient secondary language training data, to learn the discrimination between primary and secondary. The issue is resolved to some extent by considering W2V pre-trained embeddings as a feature extractor and providing an average relative improvement of 30.74\% in terms of JER. In the future, the framework can be further explored to achieve better discrimination between the languages.


\bibliographystyle{IEEEtran}

\bibliography{mybib}


\end{document}